\documentclass[12pt,onecolumn,amsmath,amssymb]{revtex4}
\usepackage{graphicx}
\begin{document}

\vspace*{4cm}

\title{Exact Solutions of the Equations of Motion of Liquid Helium with
a Charged Free Surface}

\author{N. M. Zubarev}
\email{nick@ami.uran.ru}

\affiliation{Institute of Electrophysics, Ural Branch, Russian
Academy of Sciences,\\ 106 Amundsen Street, 620016 Ekaterinburg, Russia\\
tel.: +7 343 2678776; fax: +7 343 2678794.}

\begin{abstract}
The dynamics of the development of instability of the free surface of
liquid helium, which is charged by electrons localized above it, is studied.
It is shown that, if the charge completely screens the electric field
above the surface and its magnitude is much larger then the instability
threshold, the asymptotic behavior of the system can be described by the
well-known 3D Laplacian growth equations.
The integrability of these equations in 2D geometry makes it possible to
described the evolution of the surface up to the formation of
singularities, viz., cuspidal point at which the electric field strength,
the velocity of the liquid, and the curvature of its surface assume
infinitely large values.
The exact solutions obtained for the problem of the electrocapillary wave
profile at the boundary of liquid helium indicate the tendency to a charge
in the surface topology as a result of formation of charged bubbles.
\end{abstract}

\maketitle

\section{INTRODUCTION}

It is well known \cite{1,2} that the liquid helium surface may be charged
to high values of the surface density of a negative electric charge.
This is due to the fact that, on the one hand, electrons are attracted to
the surface by weak electrostatic image forces and, on the other hand, the
liquid helium boundary is a potential barrier for electrons, which
prevents their penetration in the bulk.
An important feature of liquid helium as a dielectric with a low
polarizability is the relative weakness of the image forces, as a result
of which the mean distance between localized electrons and the surface is
much larger than the atomic spacing. Consequently, the electrons are not
bound to individual atoms of the substance and form a two-dimensional
conducting system.

\begin{figure}
\includegraphics{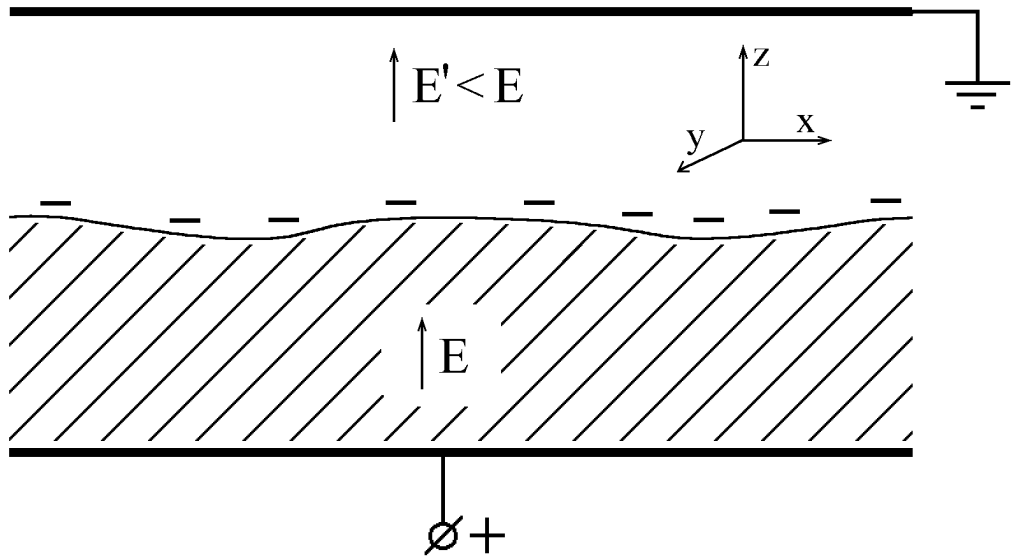}
\caption{\label{fig:fig_1}
Schematic diagram of the surface of liquid helium, charged
by electrons, in a parallel-plate capacitor.
}
\end{figure}

The ability of electrons to move freely over the surface of liquid helium
ensures the equipotential nature of this surface over characteristic
hydrodynamic times and scales.
A charged surface of a conducting liquid also possesses this property, the
only difference being that the electric field cannot penetrate into a
conducting medium, while liquid helium is not subjected to such a limitation.
This enabled Gor'kov and Chernikova \cite{3,4} to extend a number of
classical results from the theory of instability of a liquid metal surface
in an external electric field \cite{5,6,7} to the case of the charged
boundary of liquid helium (the geometry of the system is shown
schematically in Fig.~\ref{fig:fig_1}).
For example, a natural generalization of the dispersion relation for
linear waves on the surface of a conducting liquid is the following
dispersion relation for liquid helium:
\begin{equation}
\omega^2=gk+\frac{\alpha}{\rho}\,k^3-\frac{E^2+{E'}^2}{4\pi\rho}\,k^2,
\end{equation}
where $\omega$ is the frequency, $k$ is the wave number, $g$ is the
acceleration due to gravity, $\alpha$ is the surface tension, $\rho$ is
the density of the medium, and $E'$ and $E$ are the electric field
strengths above the liquid and in the bulk of it, respectively
($E=0$ for a conducting medium). It follows hence that for
$$
{E'}^2+E^2<{E_c}^2=8\pi\sqrt{g\alpha\rho}
$$
the inequality $\omega^2>0$ holds for any $k$ and, hence, small
perturbations of the surface do not build up with time.
In the case when the sum of the squares of the fields ${E'}^2+E^2$, which
plays the role of an extrinsic controlling parameter, exceeds the critical
value ${E_c}^2$, a region of wave numbers $k$ for which $\omega^2<0$ is
formed.  This corresponds to an aperiodic instability of the liquid
boundary.

The buildup of perturbations of the surface inevitably transforms the
system to a state in which its evolution is determined by nonlinear
processes. The nature of their effect can be estimated most easily in the
vicinity of the instability threshold, i.e., for a small supercriticality
$\varepsilon=(E^2+{E'}^2-{E_c}^2)/{E_c}^2$, when only perturbations with
wave numbers close to $k_0=\sqrt{g\rho/\alpha}$ increase and we can pass
to envelopes in the equations of motion.
For example, Gor'kov and Chernikova \cite{8} proved that, in the case of
2D symmetry of the problem, the complex amplitude $A(x,t)$ of perturbation
of the surface obeys the nonlinear Klein-Gordon equation
\begin{equation}
(gk_0)^{-1}A_{tt}=2\varepsilon A+k_0^{-2}A_{xx}+
\left(2S^2-5/8\right)A|A|^2,
\end{equation}
where $S=(E^2-{E'}^2)/{E_c}^2$ is the dimensionless parameter
characterizing the surface charge density.
It can be seen from Eq.~(2) that, depending on the value of parameter
$S$, the nonlinearity either saturates the instability, or, conversely,
facilitates a burst of the perturbation amplitude.
A similar conclusion can also be drawn in the general (3D) case with a
correction taking into account the fact that the nonlinearity in the first
nonvanishing order plays a destabilizing role due to the interaction of
three waves forming the hexagonal structure.
As in 2D case, cubic nonlinearities produce a stabilizing effect for small
values of $S$ \cite{9,10}. Consequently, for a low surface charge density
(when the values of $E$ and $E'$ are close), a steady-state relief of the
liquid helium boundary may be formed. In this case, the standard
perturbation theory in the small parameter, viz., the characteristic slope
of the surface, can be used for studying the structures being generated
(see \cite{11} and the literature cited therein).

The processes occurring in the supercritcal region of electric fields and
for relatively large electron surface charge screening the field above the
liquid surface to a considerable extent have not been investigated in
detail theoretically. This is due to the fact that, in these cases,
the development of instability violates the small-angle approximation.
For example, the analysis of the behavior of the charged boundary of
liquid helium by high-speed microphotography carried out by Vololdin 
{\sl et al.} \cite{12} proved that the dimples appearing on the surface are
sharpened over a finite time (the bubbles which are subsequently formed at
the tips carry the charge from the helium surface to the positive plate of
the capacitor). In view of the considerable nonlinearity of such
processes, their description requires the construction of solutions to the
fundamental equations of the electrohydrodynamics of liquid helium.

In the present work, it will be shown that, when the condition
$E\gg E'$, which corresponds to complete screening of the field above the
liquid by the surface electron charge, is satisfied along with the
condition that the electric field strength considerably exceeds the critical
value, $E\gg E_c$, the equations of motion of liquid helium have an
infinitely large number of exact analytic solutions.
Their analysis has facilitated a considerable advance in the analysis of
unsolved problems in the electrohydrodynamics of liquids with a free
surface, which are associated with the formation of singularities (cusps)
and with considerable changes in the surface geometry (formation of
bubbles).

In Section~2, the equations of a vortex-free flow of liquid helium with a
free surface charge are considered.
In the limit of a strong electric field, when the effect of the force of
gravity and capillary forces can be neglected, the approach to an analysis
of the liquid helium dynamics proposed in our earlier work \cite{13} is
developed.
This approach is based on the separation of two branches corresponding to
solutions increasing and decreasing with time in the equations of motion.
In Section~3, it is shown that the asymptotic behavior of the system is
given by the well-known equations describing the Laplacian growth in the 3D
geometry (the motion equipotential boundary with the velocity determined by
the normal derivative of the harmonic potential).
Section~4 is devoted to an analysis of the dynamics of the formation of
cuspidal dimples on the helium surface in 2D geometry, when the Laplacian
growth equations have an unlimited number of exact nontrivial solutions.
The propagation of nonlinear surface waves in the short-wave region in
which the surface pressure must be taken into account along with the
electrostatic pressure is considered in Section~5. It is shown that the
problem of the profile of a progressive electrocapillary wave at the liquid
helium boundary has exact analytic solutions similar to the Crapper
solutions for capillary waves \cite{14}. These solution are used for
obtaining a nonlinear dispersion relation for surface waves of an arbitrary
amplitude, whose analysis led to a number of conclusions concerning the
stability of the charged surface of liquid helium to finite-amplitude
perturbations and the domain of the existence of wave solutions to the
electrohydrodynamic equations. In Section~6, the simplest axisymmetric
solutions of the equations of motion, describing the pulling of the
surface into the bulk of the liquid at a constant rate, are analyzed.

\section{INITIAL EQUATIONS: THE LIMIT OF A STRONG FIELD}

Let us consider the potential motion of an ideal dielectric liquid (liquid
helium) with a free surface charged by electrons in an electric field. We
assume that, in the unperturbed state, the boundary of the liquid is a
flat horizontal surface $z=0$ and the field vector is directed along the
$z$ axis of our system of coordinates (Fig.~\ref{fig:fig_1}).
We introduce a function $\eta(x,y,t)$ specifying the deviation of the
boundary from the plane.
Then, the shape of the perturbed surface of liquid helium is described by the
equation $z=\eta(x,y,t)$. The velocity potential $\Phi$ for an
incompressible liquid satisfied the Laplace equation
\begin{equation}
\nabla^2\Phi=0,
\end{equation}
which must be supplemented with the dynamic boundary condition
\begin{equation}
\Phi_t+\frac{(\nabla\Phi)^2}{2}=\frac{E^2-(\nabla\varphi)^2}{8\pi\rho}+
\frac{\alpha}{\rho}\,\nabla_{\!\!\bot}\cdot
\frac{\nabla_{\!\!\bot}\eta}{\sqrt{1+(\nabla_{\!\!\bot}\eta)^2}}-
g\eta, \qquad z=\eta(x,y,t),
\end{equation}
where $\varphi$ is the electric potential in the liquid (we assume that
the charge completely screens the field above the helium surface). 
The first term on the right-hand side of the time-dependent Bernoulli
equation (4) is responsible for electrostatic pressure, the second 
is responsible for capillary pressure, and the third takes into account
the effect of the field of gravity. We assume that the characteristic
spatial of surface perturbations is smaller than the size of the region
occupied by the liquid. In this case, we can write
\begin{equation}
\Phi\to 0, \qquad z\to-\infty,
\end{equation}
i.e., the motion of the liquid attenuates at infinity. The time evolution
of the free surface is determined by the kinematic relation 
(the condition that the liquid does not flow through its boundary): 
\begin{equation}
\eta_t=\Phi_z-\nabla_{\!\!\bot}\eta\cdot\nabla_{\!\!\bot}\Phi,
\qquad z=\eta(x,y,t).
\end{equation}
Finally, the electric potential $\varphi$ in the absence of space charges
satisfies the Laplace equation
\begin{equation}
\nabla^2\varphi=0,
\end{equation}
which must be solved under the condition that the liquid helium boundary
is equipotential and the field is uniform at an infinitely large distance
from the surface:
\begin{equation}
\varphi=0, \qquad z=\eta(x,y,t),
\end{equation}
\begin{equation}
\varphi\to -Ez, \qquad z\to-\infty.
\end{equation}
It should be noted that, in zero electric field ($E=0$ and, hence,
$\nabla\varphi=0$) the above equations coincide with the equation of
motion for a thick layer of liquid in the field gravity.

Let the electric field strength exceed considerably its critical value 
($E\gg E_c$), and let the following relation hold for the characteristic
wavelength $\lambda$ of surface waves: 
$\alpha E^{-2}\ll\lambda\ll E^2/(g\rho)$. 
It follows from the dispersion relation (1) that, in an analysis of
small-amplitude surface perturbation, we can disregard the effect of both
the capillary forces and the force of gravity. In Secton~4, we will prove
that this statement also holds for finite-amplitude surface perturbations. 
This means that we can omit the last two terms on the right-hand side of
the boundary condition (4) and take into account the electrostatic
pressure alone.

Now we pass to the dimensionless notation, assuming that the unit of
length is equal to $\lambda$, the unit of electric field strength is
$E$, and the unit of time is $\lambda E^{-1}(4\pi\rho)^{1/2}$. In this
case, the equations of motion (3)--(9) assume the form
\begin{equation}
\nabla^2\varphi=0, \qquad
\nabla^2\Phi=0, 
\end{equation}
\begin{equation}
\Phi_t+(\nabla\Phi)^2/2+(\nabla\varphi)^2/2=1/2, \qquad z=\eta(x,y,t),
\end{equation}
\begin{equation}
\eta_t=\Phi_z-\nabla_{\!\!\bot}\eta\cdot\nabla_{\!\!\bot}\Phi,
\qquad z=\eta(x,y,t),
\end{equation}
\begin{equation}
\varphi=0, \qquad z=\eta(x,y,t),
\end{equation}
\begin{equation}
\Phi\to 0, \qquad \qquad z\to-\infty,
\end{equation}
\begin{equation}
\varphi\to -z, \qquad z\to-\infty.
\end{equation}
Let us write these equations in the form which does not contain function 
$\eta$ explicitly and introduce the perturbed harmonic potential 
$\tilde{\varphi}=\varphi+z$ attenuating at infinity 
($\tilde{\varphi}\to 0$ as $z\to-\infty$). At the boundary, we have
$\tilde{\varphi}|_{z=\eta}=\eta$. This readily leads to relations
$$
\eta_t=\left.\frac{\tilde{\varphi}_t}
{1-\tilde{\varphi}_z}\right|_{z=\eta},
\qquad
\nabla_{\!\!\bot}\eta=\left.\frac{\nabla_{\!\!\bot}\tilde{\varphi}}
{1-\tilde{\varphi}_z}\right|_{z=\eta},
$$
which allow us to eliminate $\eta$ from Eq. (12). The kinematic and
dynamic boundary conditions (11) and (12) can be transformed to
$$
\tilde{\varphi}_t-\Phi_z=-\nabla\tilde{\varphi}\cdot\nabla\Phi,
\qquad z=\eta(x,y,t),
$$
$$
\Phi_t-\tilde{\varphi}_z=-(\nabla\Phi)^2/2-(\nabla\tilde{\varphi})^2/2,
\qquad z=\eta(x,y,t).
$$
Adding and subtracting these equations, we obtain
$$
(\tilde{\varphi}+\Phi)_t-(\tilde{\varphi}+\Phi)_z=
-(\nabla(\tilde{\varphi}+\Phi))^2/2,
\qquad z=\eta(x,y,t),
$$
$$
(\tilde{\varphi}-\Phi)_t+(\tilde{\varphi}-\Phi)_z=
+(\nabla(\tilde{\varphi}-\Phi))^2/2,
\qquad z=\eta(x,y,t),
$$
i.e., the boundary conditions can be specified separately for the sum and
the difference of the harmonic potentials $\tilde{\varphi}$ and $\Phi$.
It is convenient to introduce a pair of auxiliary potentials
$$
\phi^{(\pm)}(x,y,z,t)=(\tilde{\varphi}\pm\Phi)/2.
$$
Using these potentials, we can write the equations of motion in the
following symmetric form: 
\begin{equation}
\nabla^2\phi^{(\pm)}=0,
\end{equation}
\begin{equation}
\phi^{(\pm)}_t=\pm\phi^{(\pm)}_z\mp(\nabla\phi^{(\pm)})^2,
\qquad z=\eta(x,y,t),
\end{equation}
\begin{equation}
\phi^{(\pm)}\to 0, \qquad z\to-\infty,
\end{equation}
while the shape of the liquid helium boundary is determined from the relation 
\begin{equation}
\eta=\left.(\phi^{(+)}+\phi^{(-)})\right|_{z=\eta}.
\end{equation}

Thus, the equations of motion can be split into two systems of equations
for potentials $\phi^{(+)}$ and $\phi^{(-)}$, the relation between which
is given by the implicit equation for the shape of the surface (19). 
It is important that these equations are compatible with the condition 
$\phi^{(-)}=0$ or with the condition $\phi^{(+)}=0$. In the next section,
we will show that the former condition corresponds to the solutions of the
problem whose amplitude increases with time, while the latter (which is of
no interest to us), to damped solutions.

The possibility of separating equations into individual branches is due to
the symmetries of the electrohydrodynamic equations, which can be easily
seen when the Hamilton formalism is used. Indeed, the equations of motion
(10)--(15) for a liquid with a free surface possess a Hamilton structure,
the function $\eta(x,y,t)$ and $\psi(x,y,t)=\Phi|_{z=\eta}$ being
canonically conjugate quantities \cite{15},
$$
\psi_t=-\frac{\delta H}{\delta\eta},
\qquad
\eta_t=\frac{\delta H}{\delta\psi},
$$
where the Hamiltonian $H$ coincides to within constants with the total
energy of the system: 
$$
H=K+P, \qquad
K=\int\limits_{z\leq\eta}\frac{(\nabla\Phi)^2}{2}\,d^3r,
$$
$$
P=\int\limits_{z\leq\eta}
\frac{1-(\nabla\varphi)^2}{2}\,d^3r=
-\int\limits_{z\leq\eta}
\frac{(\nabla\tilde\varphi)^2}{2}\,d^3r.
$$

It should be recalled that the harmonic potentials $\Phi$ and $\tilde\varphi$
attenuate for $z\to-\infty$ and their values on the surface are defined by
the functions $\psi$ and $\eta$, respectively. Consequently, if
$\psi=\eta$, then $\Phi=\tilde\varphi$, and the kinetic energy functional 
$K$ coincides, expect for the sing, with the potential energy functional
$P$. This allows us to write the Hamilton equations of motion using the
functional $K$ alone:
$$
\psi_t=-\frac{\delta K}{\delta\eta}
+\left.\left(\frac{\delta K}{\delta\eta}
+\frac{\delta K}{\delta\psi}\right)\right|_{\psi=\eta},
\qquad
\eta_t=\frac{\delta K}{\delta\psi}.
$$
It can be seen that, if we set $\psi=\eta$ in these equations, they will
coincide. This means that the condition $\psi=\eta$ or (which is the
same) the condition $\phi^{(-)}=0$ is compatible with the equations of
motion for liquid helium. Similarly, we can prove that the Hamilton
equations  coincide for $\psi=-\eta$, which corresponds to the condition 
$\phi^{(+)}=0$. It should also be noted that the equations describing the
evolution of the system on the branches $\phi^{(+)}=0$ and $\phi^{(-)}=0$ 
coincide except for the substitution $t\to-t$, which is associated with
the time reversibility in the Hamilton equations of motion. In this case,
the conditions $\phi^{(\pm)}=0$ single out the solutions of the problem
for which $H$ is equal to zero.

\section{INCREASING BRANCH: STABILITY}

In the linear approximation whose applicability is limited by the
condition of the smallness of the slopes of the surface 
$|\nabla_{\!\!\bot}\eta|\ll 1$, the boundary conditions (17) assume the
form 
$$
\phi^{(\pm)}_t=\pm\phi^{(\pm)}_z, \qquad z=0,
$$
and Eqs. (16)--(19) split into two independent systems. The dispersion
relations for these systems can be found by substituting potentials in the
form 
$\phi^{(\pm)}\sim e^{kz+i{\bf k}{\bf r}_{\!\bot}-i\omega t}$. This gives 
$$
\omega^{(\pm)}=\pm ik
$$
(the same result follows directly from the dispersion relation (1)
considered in the strong field limit). It can be seen that, for one
branch, small periodic perturbations of the surface increase exponentially
with the characteristic times $k^{-1}$, while, for the other branch, these
perturbations attenuate. In this case, for large periods of time, we can
assume that $\phi^{(-)}=0$ and consider only equations for potential 
$\phi^{(+)}$. Let us prove that this statement is also valid in the
general case, when the evolution of the surface is described by nonlinear
equations (16)--(19). 

We assume that, in the nonlinear equations of motion (16)--(19), 
$$
\phi^{(+)}=\varphi+z, \qquad \phi^{(-)}=0,
$$
which, in accordance with the results of linear analysis, isolates the
solutions increasing with time. Passing to the moving fame of reference 
$\{x,y,z'\}=\{x,y,z-t\}$ in which the plane unperturbed surface of the
liquid moves downwards (i.e., in the direction opposite to the $z'$ axis)
at a constant velocity, after simple transformations, we obtain
\begin{equation}
\nabla^2\varphi=0, 
\end{equation}
\begin{equation}
\eta'_t=\partial_n\varphi\,\sqrt{1+(\nabla_{\!\!\bot}\eta')^2}, 
\qquad z'=\eta'(x,y,t).
\end{equation}
\begin{equation}
\varphi=0, \qquad z'=\eta'(x,y,t)
\end{equation}
\begin{equation}
\varphi\to -z', \qquad z'\to-\infty,
\end{equation}
where $\eta'(x,y,t)=\eta-t$ and $\partial_n$ denotes the derivative along
the normal to the boundary of the liquid. These equations define
explicitly the motion of the free charged surface of liquid helium 
$z'=\eta'(x,y,t)$. They coincide with the equations describing the
so-called Laplacian growth, viz., the motion of the phase boundary with a
velocity directly proportional to the normal derivative of a certain
harmonic scalar field ($\varphi$ in our case). Depending on the chosen
frame of reference, this field may have the meaning of temperature
(Stefan's problem in the quasi-stationary limit), electrostatic potential
(electrolytic deposition), or pressure (flow through a porous medium). 

Let us prove that the solutions of Eqs. (10)--(15) corresponding to system 
(20)--(23) are stable to small perturbations of potential $\phi^{(-)}$. 
It should be noted that the motion of the liquid boundary described by
Eqs. (20)--(23) is always directed inwardly; this is associated with the
principle of the extremum for harmonic functions. Let function $\eta'$ at
the initial instant $t=0$ be a single-valued function of variables $x$ and
$y$. In this case, for $t>0$, the following inequality holds: 
$\eta'(x,y,t)\leq\eta'(x,y,0)$. In the original notation, we have
\begin{equation}
\eta(x,y,t)\leq\eta(x,y,0)+t
\end{equation}
for any $x$ and $y$. This inequality remains valid for small perturbations
of $\phi^{(-)}$ also, when the effect of potential $\phi^{(-)}$ in
relation (19) can be disregarded as compared to the effect of potential 
$\phi^{(+)}$, and the motion of the boundary is described by the same
Eqs. (20)--(23). 

As regards the evolution of potential $\phi^{(-)}$, it is described, for
small $|\nabla\phi^{(-)}|$, by Eqs.~(16)--(18), where it is sufficient to
consider the condition (17) at the boundary in the linear approximation:
$$
\phi^{(-)}_t=-\phi^{(-)}_z, \qquad z=\eta(x,y,t).
$$
Let us suppose that, at the initial instant $t=0$, the potential
distribution is described by the following expression:
$$
\phi^{(-)}|_{t=0}=\phi_0(x,y,z),
$$
where $\phi_0$ is a certain function which is harmonic for
$z\leq\eta(x,y,0)$ and attenuating for $z\to-\infty$. In this case, the
temporal dynamics of potential $\phi^{(-)}$ is described by the expression
$$
\phi^{(-)}=\phi_0(x,y,z-t).
$$
It can be seen from this expression that the singularities of the function
$\phi^{(-)}$ are displaced in the direction of the $z$ axis and can exist
only in the region
\begin{equation}
z>\eta(x,y,0)+t.
\end{equation}
A comparison of this inequality with (24) shows that the singularities of
potential $\phi^{(-)}$ do not approach the liquid helium boundary
$z=\eta(x,y,t)$ and, hence, the value of the potential at the surface does
not increase with time. It should be noted that, otherwise, the solutions
obtained by us for $\phi^{(-)}$ would be inapplicable. 

In view of incompressibility of the liquid, the level of its surface (the
value of function $\eta$ averaged over the spatial variables) is not
displaced. On the other hand, the boundary of the region defined by
inequality (25) and averaged over $x$ and $y$, in which singularities of
the function $\phi^{(-)}$ occurs, moves upwards at a constant velocity.
This means that the singularity moves away from the surface of liquid
helium and the perturbation of $\phi^{(-)}$ relaxes to zero. 

Thus, we have proved that, as $t\to\infty$, we have 
$$
\varphi(x,y,z,t)+z\to\Phi(x,y,z,t), 
$$
and Eqs. (20)--(23) describe the asymptotic behavior of liquid helium with
a charged surface in a strong electric field.

\section{SOLUTIONS OF 2D EQUATIONS OF MOTION}

In the previous section, we proved that the analysis of the 3D potential
motion of liquid helium in a strong electric field can be reduced to
analysis of Eqs.~(20)--(23) describing the three-dimensional Laplacian
growth. The exact solvability of these equations in the 2D geometry will
allow us to effectively study the dynamics of the development of
instability of the charged surface of a liquid, including the formation of
singularities in it.

We assume that, in the system of equations (20)--(23), all quantities are
independent of variable $y$ (variable $y'$). We introduce the function
$w=v-i\varphi$ of the complex argument $Z=x+iz'$, which is analytic for  
$z'\leq\eta'(x,t)$ (this is the co-called complex potential of the field
correct to a constant factor). Here, $v$ is a function harmonically
conjugate to $\varphi$ and such that the condition $v=\mbox{const}$
defines the electric field lines in the medium. Clearly,
$w\to Z$ as $Z\to x-i\infty$. 

It is convenient for the subsequent analysis to pass to a system of
coordinates in which the role of the independent variable is played by
quantity $w$ and the role of the unknown function is played by function
$Z$ which is analytic in the lower half-plane of the complex variable $w$
(i.e., for $\varphi>0$). It follows from condition (23) that the following
condition holds at infinity: 
\begin{equation}
Z\to w, \qquad w\to v-i\infty.
\end{equation}
We can also obtain the condition for $Z$ at the boundary $\varphi=0$ of
the half-plane. The profile of the liquid helium surface can be specified
by the parametric relations  
$$
z'=z'(v,t)=\eta'(x(v,t),t), \qquad x=x(v,t).
$$
Using these relations, we can easily express the normal velocity of the
surface and the electric field strength appearing in formulas (21) in 
terms of the functions $z'(v,t)$ and $x(v,t)$: 
$$
\frac{\eta'_t}{\sqrt{1+{\eta'_{x}}^2}}=
\frac{z'_t x_v-x_t z'_v}
{\sqrt{{z'}^2_v+x_v^2}},
\qquad
\partial_n\varphi=-\frac{1}{\sqrt{{z'}^2_v+x_v^2}}.
$$
Substituting these relations into the condition (21) at the surface, we
obtain 
$$
z'_t x_v-x_t z'_v=-1,
$$
or, which is the same,
\begin{equation}
\mbox{Im}(Z^*_t Z_w)=1, \qquad w=v.
\end{equation}

Thus, we arrive at the problem of determining the function $Z$, which is
analytic in the lower half-plane of the complex variable $w$ and satisfies
conditions (26) and (27). The nonlinear condition (27) is the so-called
Laplacian growth equation which is widely used for describing the 2D
motion of the boundary between two liquids with noticeably different
viscosities \cite{16,17}, the evolution of the free surface of a liquid in
the field of gravity \cite{18,19}, and so on. The Laplacian growth equation
is integrable in the sense that it has an infinitely large number of
particular solutions of the form \cite{20} 
\begin{equation}
Z(w)=w-it-i\sum_{n=1}^N a_n\ln\left(w-w_n(t)\right)
+i\left(\sum_{n=1}^N a_n\right)\ln\left(w-w_0(t)\right).
\end{equation}
Here, $a_n$ are complex constants, and the functions of time $w_n$ satisfy
the condition $\mbox{Im}(w_n)>0$ (singularities of the function $Z$ can
only be in the upper half-plane of of the complex variable $w$). The last
term in expression (28) was supplemented to ensure the fulfillment of
condition (26) and, hence, the condition of localization of the
perturbation of the surface in a certain region: $\eta\to 0$ for
$|x|\to\infty$. We can set $\mbox{Im}(w_0)\gg\mbox{Im}(w_n)$; in this
case, the effect of this term on the evolution of the surface is
negligibly small.

Substituting expression (28) into Eq.~(27) and decomposing the obtained
expression into simple fractions, we obtain a system of $N$ ordinary
differential equations for $w_n(t)$: 
$$
\dot{w}_n+i+i\sum_{m=1}^N
a^*_m\,\frac{\dot{w}_n-\dot{w}^*_m}{w_n-w^*_m}=0.
$$
Integration with respect to $t$ leads to the following $N$ transcendental
equations: 
$$
w_n+it+i\sum_{m=1}^Na^*_m\ln\left(w_n-w^*_m\right)=C_n,
$$
where $C_n$ are arbitrary complex constants.

Let us consider the simplest solutions of this type, which correspond to 
$N=1$, $\mbox{Re}(w_1)=0$ and $a_1=\pm 1$:
\begin{equation}
Z(w)=w-it\mp i\ln(w-iq(t)),
\end{equation}
\begin{equation}
q(t)\pm\ln q(t)=1+t_c-t, 
\end{equation}
where $q=\mbox{Im}(w_1)$ and $t_c$ is a real constant. The form of a
solitary perturbation corresponding to Eqs. (29) and (30) is specified by
the parametric expressions 
$$
z(v,t)=z'(v,t)+t=\mp\ln\sqrt{v^2+q^2(t)},
$$
$$
x(v,t)=v\pm\arctan\left(v/q(t)\right).
$$

\begin{figure}
\includegraphics{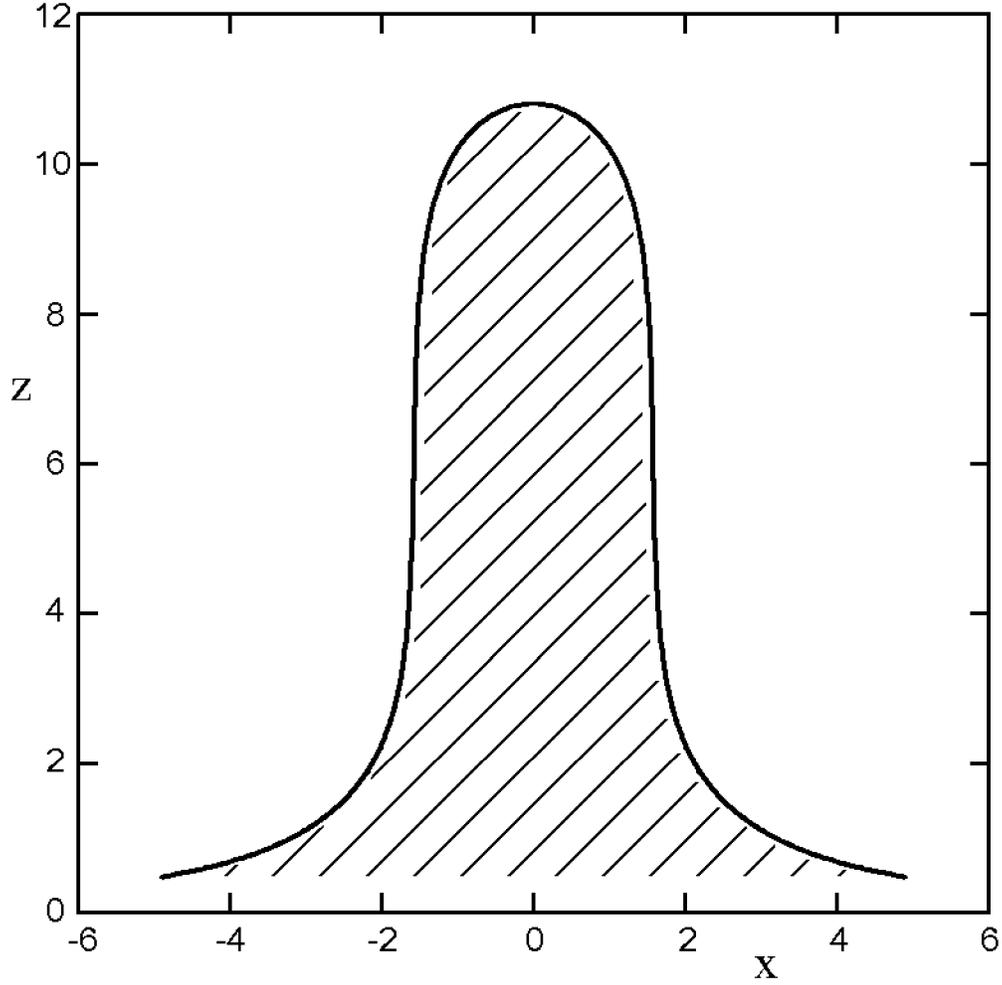}
\caption{\label{fig:fig_2}
The profile of the liquid helium surface, corresponding to
the "one-finger" solution of the Laplacian growth equation; $a_1=1$, 
$q=10^{-4}$, and $w_0=i5$.
}
\end{figure}

Let us suppose that $a_1=+1$ and we are dealing with a solitary perturbation
of the surface, which is directed "upwards". It can be seen from Eq. (30)
that for large values of $t$, the quantity $q\sim e^{-t}$ and, hence, the
surface perturbation amplitude increases linearly with time: $z|_{v=0}\to
t$ as $t\to\infty$. This is the "one-finger" solution of the Laplacian growth
equation (see Fig.~\ref{fig:fig_2}). It can easily be proved that similar solutions are 
possible in the 3D case also. It can be seen from Eqs.~(20)--(23) describing
the three-dimensional Laplacian growth that, if the surface initially
contains a region in which the field strength $\partial_n\varphi$ is small
(e.g., in the vicinity of the apex of a 3D fingerlike perturbation of the
surface), its velocity in the coordinates $\{x,y,z'\}$ is also small.
In the laboratory reference frame, this corresponds to a jet flowing at a
constant velocity in the direction of the $z$ axis.

\begin{figure}
\includegraphics{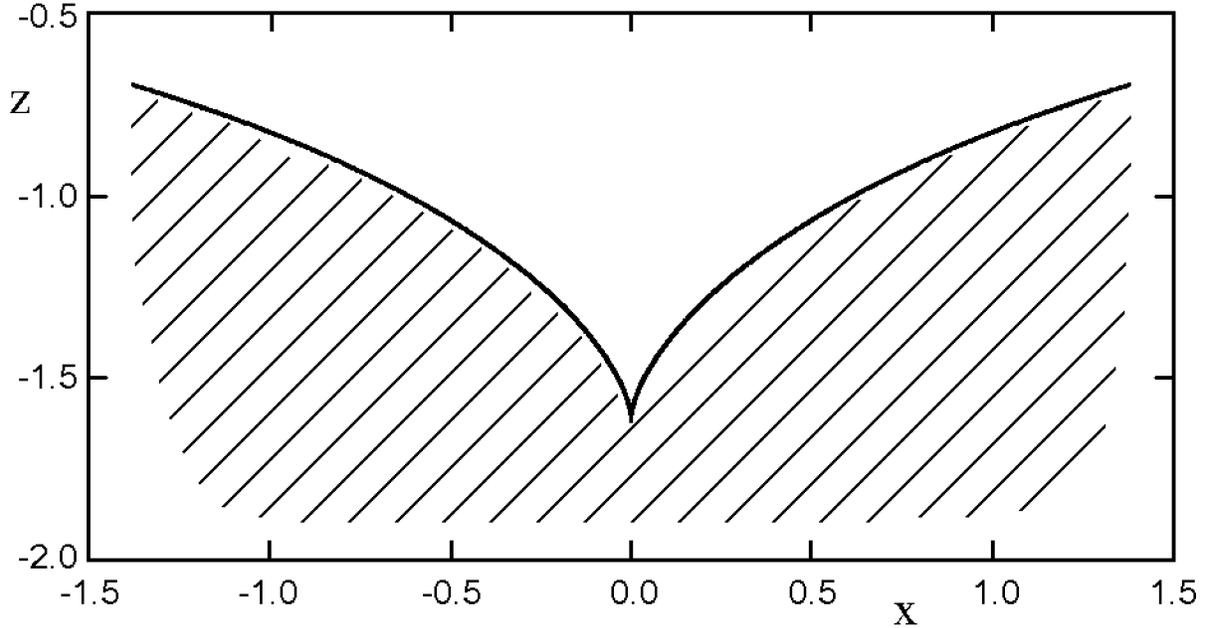}
\caption{\label{fig:fig_3}
The profile of the liquid helium surface at the instant of
formation of a singularity (cusp); $a_1=-1$, $q=0.8$, and $w_0=i4$.
}
\end{figure}

Let us now consider a solitary perturbation of the surface, which is directed
"downwards" ($a_1=-1$, $q(t)\geq 1$). This solution exists only during a
finite period of time, leading to the formation of a singularity on the
liquid surface, viz., cuspidal point of the first kind (Fig.~\ref{fig:fig_3}), at
instant $t=t_c$. Indeed, expanding $z$ and $x$ into power series in $v$
and $\tau$ taking into account the fact that the function $q(t)$ in the
vicinity of the $t_c$ satisfies the relation 
$$
q(t)\approx 1+\sqrt{2\tau}, \qquad \tau=t_c-t,
$$
we obtain the following expressions in the main order:
\begin{equation}
z=v^2/2+\sqrt{2\tau}, \qquad
x=v^3/3+v\sqrt{2\tau}.
\end{equation}
It can be seen that, at instant $\tau=0$ (i.e., for $t=t_c$), the shape of
the surface in the vicinity of a singularity point is defined by the
relation $2z=|3x|^{2/3}$, which corresponds to a cusp \footnote{On the
charged surface of a conducting liquid, for which $E=0$ and $E'\not=0$, in
the limit of a strong field, weak root singularities of the type $z\sim
|x|^{3/2}$ are formed, for which curvature is equal to infinity, while the
surface itself remains smooth \cite{21,22}.}. It was indicated in
\cite{17,23} that the singularities of $z^3\sim x^2$ are general-position
singularities for processes described by the Laplacian growth equation. 
Similar solutions of the equations of motion for liquid helium with the
charged boundary reflect the experimentally observed tendency \cite{12,24}
to the emergence of dimples on the surface, which become sharpened over a
finite time. From the mathematical point of view, the emergence of
singularity on the liquid surface is associated with vanishing of the
Jacobian of the transformation $\{x,z'\}\to\{v,\varphi\}$ for
$\varphi=v=\tau=0$. At a cusp, the electric field strength increases
indefinitely along with the velocity of the surface over a finite time
interval: 
$$
|\nabla\varphi|\sim\left.x_v^{-1}\right|_{v=0}\sim\tau^{-1/2},
\qquad
|\nabla\Phi|\sim\left.z_t\right|_{v=0}\sim\tau^{-1/2}. 
$$

It is important to note that the singular solution of the problem
described by expressions (31) is also valid in the case when the field
above the surface is not screened completely; i.e., the condition $E'\ll
E$ does not hold. As a matter of fact, in the vicinity of a singularity,
the condition of the smallness of the field above the surface as compared
to the field in the bulk of the liquid naturally holds. In addition, the
condition $\lambda\ll E^2/(g\rho)$ is not necessary. This is due to the
fact that the amplitude of surface perturbations remains finite, and the
effect of the gravity forces is always negligibly small in the vicinity of
the cusp. 

Let us now consider the capillary effects. The surface and the
electrostatic pressure in the vicinity of a singularity can be estimated
easily: 
$$
P_S\sim\alpha\eta_{xx}\sim\alpha\rho^{1/2}E^{-1}\tau^{-1},
\qquad
P_E\sim(\nabla\varphi)^2\sim\lambda\rho^{1/2}E\tau^{-1}.
$$ 
Here, we have returned to the dimensional notation. It can be seen from
these expression that, when the condition $\lambda\gg\alpha E^{-2}$ is
satisfied, the capillary forces are small as compared to electrostatic
forces and, hence, can be disregarded at the stage of formation of cusps. 
This is the only necessary condition of the applicability of the Laplacian
growth equation and its solutions (31) in the vicinity of singularities.

\section{ELECTROCAPILLARY WAVES}

Let us consider the case when the characteristic length of the surface
waves is comparable with the value of $\alpha E^{-2}$ and the capillary
effects must be taken into consideration. We assume that condition $E\gg
E_c$ is satisfied; in this case, the effect of the force of gravity can be
neglected. The dispersion relation (1) for electrocapillary waves at the
charged boundary of liquid helium for $E'=0$ in the dimensionless notation
introduced in Secton~2 assumes the form
\begin{equation}
\omega^2(k)=k^3-k^2,
\end{equation}
where the value of $\lambda=4\pi\alpha E^{-2}$ is taken for unit length. 
It can be seen from Eq. (32) that $\omega^2<0$ for $k<1$ and, hence,
aperiodic electrohydrodynamic instability of the liquid surface develops.
If, however, the condition $k>1$ holds, the frequency $\omega$ is
real-valued, which corresponds to the propagation of linear dispersive
waves. 

The approach to the study of the evolution of a charged liquid surface
based on the analysis of relation (32) is obviously applicable only in the
case of small-amplitude perturbations of the boundary: $A\ll k^{-1}$. For
finite-amplitude waves, the nonlinear effect may consist in the dependence
of the dispersion relation on $A$ (see, for example, \cite{25}): 
$$
\omega=\omega(k,A).
$$
The amplitude dependence of frequency is usually sought in the form of a
power series in $A$ (Stokes expansion), which limits the analysis to the
weak-nonlinearity limit. Let us prove that, for electrocapillary waves, an
exact solution to the nonlinear dispersion relation can be found. 

The equations describing a progressive wave (whose profile does not change
in the reference frame attached to the wave) can be obtained from the
electrohydrodynamic equations (3)--(9) with the help of the following
substitutions: 
$$
\varphi=\varphi(x',z), \qquad
\Phi=\Phi'(x',z)+Cx', \qquad
\eta=\eta(x'),
$$
where $x'=x-Ct$ and constant $C$ has the meaning of the velocity of a wave
moving in the direction of the $x$ axis. This gives
\begin{equation}
\Phi'_{x'x'}+\Phi'_{zz}=0, 
\end{equation}
\begin{equation}
\varphi_{x'x'}+\varphi_{zz}=0,
\end{equation}
\begin{equation}
\frac{{\Phi'_{x'}}^2+{\Phi'_z}^2-C^2}{2}+
\frac{{\varphi_{x'}}^2+{\varphi_z}^2-1}{2}=
\frac{\eta_{x'x'}}{\left(1+{\eta_{x'}}^2\right)^{3/2}},
\qquad z=\eta(x'),
\end{equation}
\begin{equation}
\Phi'_z=\eta_{x'}\Phi'_{x'}, \qquad z=\eta(x'),
\end{equation}
\begin{equation}
\varphi=0, \qquad z=\eta(x'),
\end{equation}
\begin{equation}
\Phi'\to-Cx', \qquad z\to-\infty,
\end{equation}
\begin{equation}
\varphi\to-z, \qquad z\to-\infty.
\end{equation}
These equations can be simplified by introducing the function of current
$\Psi(x',z)$, which is harmonically conjugate to potential $\Phi'$: 
$$
\Psi_{x'}=-\Phi'_z, \qquad \Psi_{z}=\Phi'_{x'}.
$$
This function satisfies the Laplace equation
\begin{equation}
\Psi_{x'x'}+\Psi_{zz}=0
\end{equation}
with the boundary conditions
\begin{equation}
\Psi=0, \qquad z=\eta(x'),
\end{equation}
\begin{equation}
\Psi\to-Cz, \qquad z\to-\infty,
\end{equation}
which follow from relations (36) and (38). It can easily be seen that
Eqs.~(40)--(42) coincide with the Eqs.~(34), (37) and (39) for the electric 
potential. Consequently, the following functional relation exists: 
$$
\Psi=C\varphi.
$$
Using this relation, we can considerably simplify the Bernoulli equation
(35), which assume the form 
\begin{equation}
\frac{C^2+1}{2}\,\left({\varphi_{x'}}^2+{\varphi_z}^2-1\right)=
\frac{\eta_{x'x'}}{\left(1+{\eta_{x'}}^2\right)^{3/2}},
\qquad z=\eta(x').
\end{equation}
In combination with relations (34), (37), and (39), this condition
completely defines the shape of a wave propagating in the coordinate
system $\{x',z\}$.

Equations (34), (37), (39) and (43) coincide except for constant factors
with the equations describing the shape of a progressive capillary wave
\cite{14} and an equilibrium configuration of the charged surface of the
liquid metal \cite{26}. These equations have exact periodic solutions for
which the boundary of the liquid is defined by the parametric expressions
\begin{equation}
z=\frac{4k^{-2}}{2(C^2+1)^{-1}+A\cos(kp)}+z_0,
\end{equation}
\begin{equation}
x'=p-\frac{2Ak^{-1}\sin(kp)}{2(C^2+1)^{-1}+A\cos(kp)}+x'_0,
\end{equation}
where $z_0$ and $x'_0$ are constants, $p$ is a parameter (the value of $p$
changes over a period by $2\pi/k$), and the quantity $A$ has the meaning
of the amplitude of a surface perturbation; i.e.,
$A=(z_{\max}-z_{\min})/2$. The dependence of $A$ on $k$ and $C$ is
specified by the relation 
\begin{equation}
A=\left[\frac{4}{(C^2+1)^2}-\frac{4}{k^2}\right]^{1/2}.
\end{equation}
It was mentioned in \cite{14} that solutions (44) and (45) exist only for 
$1\leq k/(C^2+1)\leq\gamma$, where $\gamma\approx1.52$. 

\begin{figure}
\includegraphics{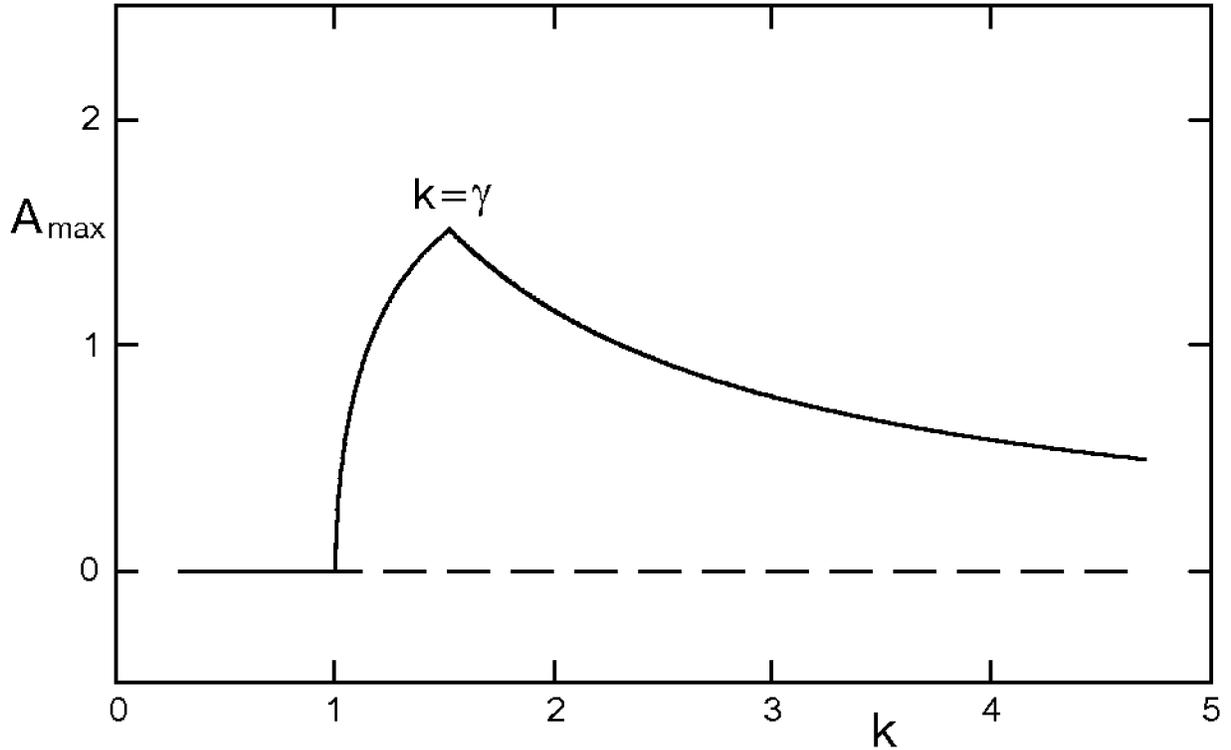}
\caption{\label{fig:fig_4}
The maximum value of amplitude $A_{\max}$ of an
electrocapillary wave on the charged surface of liquid helium as a
function of the wave number $k$. For $k<\gamma$, the peak corresponds to the
value of $\omega=0$, while, for $k>\gamma$, the frequency differs from
zero.
}
\end{figure}

Considering that $C$ is the phase velocity of the wave, we set
$C=\omega/k$ in relation (46). Solving the obtained equation for frequency
$\omega$, we arrive the exact nonlinear dispersion relation 
\begin{equation}
\omega^2(k,A)=\frac{k^3}{\sqrt{1+A^2k^2/4}}-k^2,
\end{equation}
and the conditions of its applicability 
\begin{equation}
k^3\gamma^{-1}\leq\omega^2-k^2\leq k^3.
\end{equation}
It can be seen that, in the limit of infinitely small amplitudes ($A\to
0$), expression (47) is transformed into the linear dispersion relation (32). 
Let us consider the consequences of this nonlinearity. It can be seen from
relation (47) that, for a fixed wave number $k\geq 1$, the maximum value
of the surface perturbation amplitude $A_{\max}(k)$ corresponds to the
minimum possible value of $\omega^2$. It follows from conditions (48) that,
for $1\leq k\leq\gamma$, the value of $\omega^2_{\min}=0$, which
corresponds to a wave with zero velocity. In this case, expressions (44)
and (45) define the solution of the problem on the steady-state profile of
the charged surface of the liquid helium. For $k>\gamma k_1$, the
amplitude has the maximum value for electrocapillary waves propagating at the
velocity 
$C=\sqrt{k\gamma^{-1}-1}$; in this case, $\omega^2_{\min}=k^3\gamma^{-1}-k^2$.
This gives 
$$
A_{\max}(k)=\left\{ \begin{array}{ccc}
0,\quad & 0\leq k<1 \\ 
2\sqrt{1-k^{-2}},\quad & 1\leq k\leq\gamma \\ 
2k^{-1}\sqrt{\gamma^2-1},\quad & k>\gamma
\end{array}\right.
$$
(the curve describing this dependence is presented in Fig.~\ref{fig:fig_4}). If the
amplitude exceeds this value, expressions (44) and (45) describe a
self-intersecting surface, which cannot be realized from the physical
considerations, or $\omega^2<0$, which corresponds to incorrectly
formulated problem in the context wave propagation. This leads to the
assumption that the condition $A(k)>A_{\max}(k)$ is the criterion of hard
excitation of electrohydrodynamic instability of a plane charged suface of
liquid helium, which generalizes the simplest linear instability criterion
$k<1$ to the case of finite-amplitude perturbation.

\begin{figure}
\includegraphics{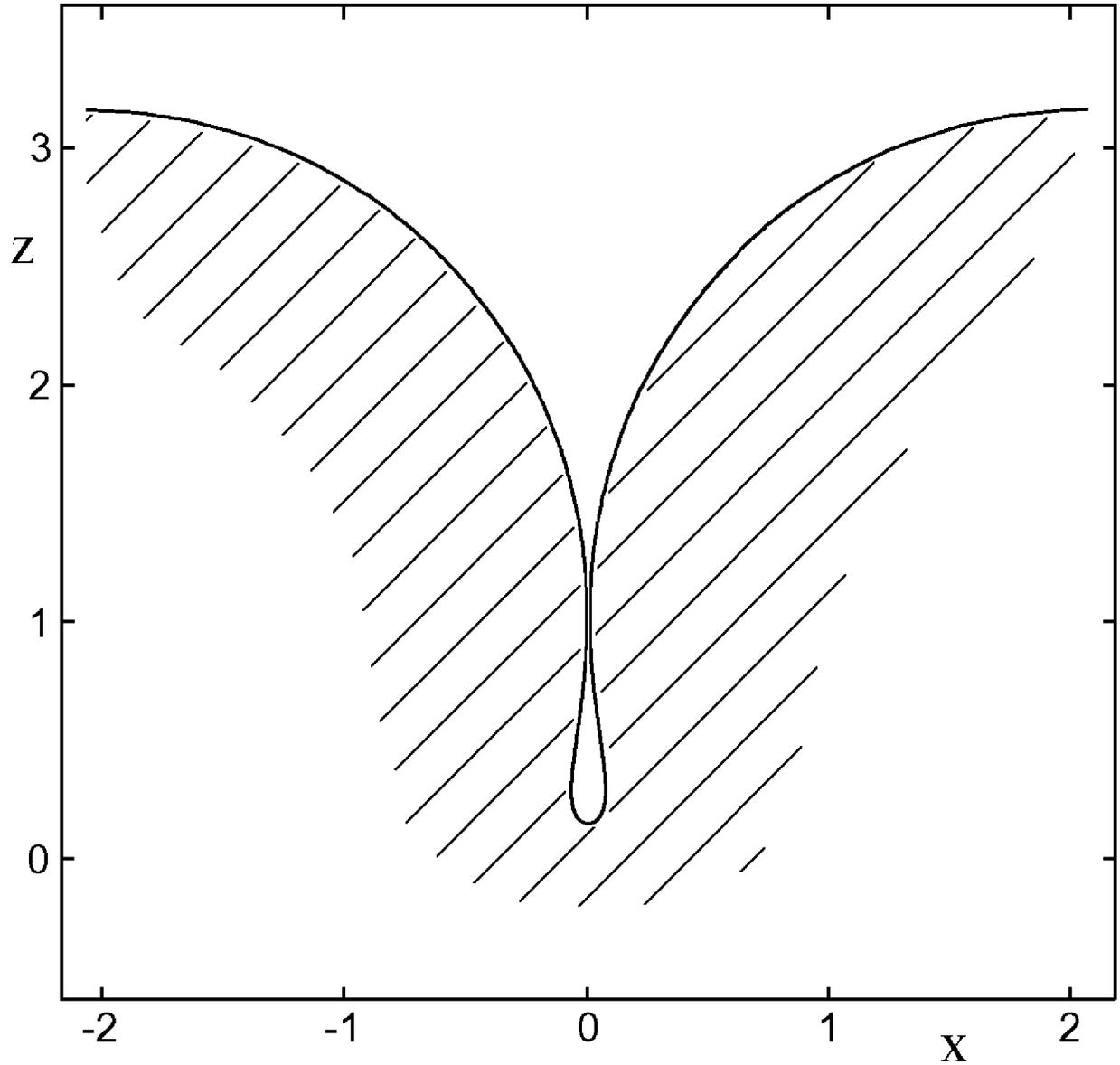}
\caption{\label{fig:fig_5}
A period of the steady-state profile of a charged surface of
liquid helium for $k=\gamma$. For this value of the wave number, the
amplitude of the electrocapillary wave attains its maximum value.
}
\end{figure}

It should be noted that peak of the function $A_{\max}(k)$ corresponds to
$k=\gamma$. The shape of the liquid surface corresponding to this value of
the wave number is depicted in Fig.~\ref{fig:fig_5}. It can be seen that the liquid
acquires cavities. Such solutions reflect the tendency to the formation of
charged bubbles (referred to as bubblons in the experimental work
\cite{12}) on cuspidal dimples of the liquid helium boundary. The main
mechanism of departure of electrons from the surface is associated with
the generation of such bubbles.

\section{AXISYMMETRIC SOLUTIONS}

Let us consider the evolution of the charged surface of liquid helium in
an important case of the axial symmetry of the problem. The equations of
motion (20)--(23) corresponding to the increasing branch of the solutions
to system (10)--(15) in the cylindrical coordinaties $\{r,z'\}=\{r,z-t\}$
assume the form 
$$  
\varphi_{rr}+r^{-1}\varphi_r+\varphi_{z'z'}=0,
$$
$$ 
\eta'_t=-(\varphi_r^2+\varphi_{z'}^2)^{\frac{1}{2}}
(1+{\eta'}_r^2)^{\frac{1}{2}}, 
\qquad z'=\eta'(r,t),
$$
$$ 
\varphi=0,
\qquad z'=\eta'(r,t),
$$
$$ 
\varphi\to -z', \qquad z'\to-\infty.
$$
Here, $r=\sqrt{x^2+y^2}$ and we have taken into account the fact that
$\partial_n \varphi=-|\nabla\varphi|$ at the equipotential boundary in
condition (21). 

At essentially nonlinear stages of the formation of a dimple on the
surface of a liquid, we can assume that the electric field in the region
of a large curvature of the surface is much stronger than the external
field, $|\nabla\varphi|\gg 1$. In this case, the dynamics of the boundary 
$\eta'=\eta-t$ is completely determined by the intrinsic field rapidly
attenuating with increasing distance, which allows us to use the condition
$|\nabla\varphi|\to 0$ for $z\to-\infty$ instead of the condition of field
uniformity. We will also assume that the velocity of the liquid surface is
considerably higher than the velocity of the origin in the laboratory
reference frame (i.e., $|\eta_t|\gg 1$). In this case, we can substitute
$\eta$ for $\eta'$ and $z$ for $z'$. This gives 
\begin{equation} 
\varphi_{rr}+r^{-1}\varphi_r+\varphi_{zz}=0, \qquad z<\eta(r,t),
\end{equation}
\begin{equation}
\varphi_t=-{\varphi_r}^2-{\varphi_{z}}^2, 
\qquad z=\eta(r,t),
\end{equation}
\begin{equation}
\varphi=0,
\qquad z=\eta(r,t),
\end{equation}
\begin{equation}
{\varphi_r}^2+{\varphi_z}^2\to 0, \qquad r^2+z^2\to\infty.
\end{equation}
In relation (50), we have used the following conditions at the boundary of
the liquid: 
$$
\eta_t=-\varphi_t/\varphi_{z}, \qquad 
\eta_r=-\varphi_r/\varphi_{z}.
$$
The conditions of the applicability of this approximation will be
considered at the end of this section. 

A particular solution of Eqs.~(49)--(52) can be obtain by using a
substitution similar to that used in \cite{27} for constructing the
axisymmetric solutions of the Stefan problem: 
\begin{equation}
\varphi(r,z,t)=f(u(r,z,t)),
\end{equation}
\begin{equation}
u(r,z,t)=-z-Vt+\sqrt{r^2+(z+Vt)^2},
\end{equation}
where the constant $V$ has the meaning of the inward-directed velocity of
the liquid surface. It can easily be seen that the equipotential surfaces
corresponding to Eqs. (53) and (54) form a family of confocal paraboloids
of revolution:  
\begin{equation}
r^2=2u(z+Vt)+u^2
\end{equation}
with the focus at the point $r=0$ and $z=-Vt$.

Substituting expressions (53) and (54) into Eq. (49), we arrive at the
following ordinary differential equation: 
\begin{equation}
uf_{uu}+f_{u}=0.
\end{equation}
It follows from Eqs. (50) and (51) that the boundary conditions to this
equation have the form 
\begin{equation}
f_u(u_0)=V/2, \qquad f(u_0)=0. 
\end{equation}
Here, $u_0$ is the value of parameter $u$ at the surface of the liquid.
Henceforth, we will use the quantity $K=1/u_0$ which, in accordance with
Eq. (55), defines the curvature of the liquid surface at the symmetry
axis. Solving Eqs. (56) and (57), we obtain 
\begin{equation}
f(u)=V\ln(Ku)/(2K),
\end{equation}
which, together with Eqs. (53) and (54), describes the time evolution of
the electric potential. It should be noted that condition (52) is
naturally satisfied. The shape of the surface for the given exact solution
of the equations of motion (49)--(52) is defined by the relation 
\begin{equation}
\eta(r,t)=Kr^2/2-Vt-(2K)^{-1},
\end{equation}
which corresponds to needle-shaped dimple pulled into the bulk of the
liquid velocity $V$. Such a geometry of the surface perturbation can be
regarded as the simplest (paraboloidal) approximation of the shape of the
liquid boundary at essentially nonlinear stages of the development of
instability of the charged boundary of the liquid.

It should be recalled that the applicability of approximation (49)--(52)
of the initial system (20)--(23) is limited by the conditions
$|\eta_t|\gg 1$ and $|\nabla\varphi|\gg 1$.
Since $\eta_t=-V$ in solutions (59) for any $r$ and $t$, the first
condition is reduced to the inequality $V\gg 1$ (in the dimensional
notation, $V\gg E\sqrt{4\pi\rho}$). As regards the second condition, we
can find the characteristic size $D$ of the region in which the electric
field created by a charged paraboloidal surface exceeds the external
field. It follows from relations (53), (54) and (58) that the field
distribution in the liquid is described by the relation
$$
|\nabla\varphi|=\frac{V}{K\sqrt{2Ru}}.
$$
Here, $R=\sqrt{r^2+(z+Vt)^2}$ is the distance to the focus of the
paraboloid; i.e., the field attains it maximum value equal to $V$ at the
point $r=0$ and $z=-Vt-(2K)^{-1}$. Since the field strength generally
decreases in proportion to $R^{-1}$ with increasing distance from the
focus, the scale of $D$ can be estimated as $D\sim V/K$. It should be
noted that such a conclusion makes sense only if the value of $D$ is much
larger than the radius of the curvature $K^{-1}$ of the surface perturbation.
This requirement again leads us to the inequality $V\gg 1$.

Thus, we have obtained partial axisymmetric solutions to the equations of
motion of liquid helium with a charged surface, which describe the
evolution of a localized perturbation of the surface with a considerable
curvature, and have determined the conditions of their applicability.
However, the obtained solutions should not be regarded as general-position
solutions. In all probability, solutions of the burst type, for which the
surface becomes indefinitely cuspidate over a finite time interval, will
dominate as in the 2D case.

\section{CONCLUDING REMARKS}

In the absence of a surface charge, the equations of electrohydrodynamics of
liquid helium considered by us are transformed into the well-known
equations of a vortex-free flow of an incompressible liquid with a free
boundary. These equations are extremely difficult to analyze, and the
methods for the solution have not been developed at present. In this work,
we succeeded in proving that the inclusion of the electrostatic pressure
does not complicate the analysis of these equations. On the contrary, the
emergence of an additional term in the dynamic boundary condition
introduces a certain symmetry into the equations so that they become
compatible with the conditions $\varphi+z=\pm\Phi$. The emerging
functional relation between the potentials of velocity and of electric
field makes it possible to reduce by half the number of equations required
for describing the motion of the surface and, in the long run, to find a
wide class of exact solutions of the equations of motion of liquid helium
with the boundary charged by electrons. It is important that the solutions
obtained by us are not limited by the condition of smallness of surface
perturbations; they describe the evolution of the liquid boudary up to the
formation of cuspidal points in it.  

The dynamics of the formation of singularities in the case when the
characteristic scale $\lambda$ of surface perturbations is comparable with
the value of $\alpha E^{-2}$ and the capillary effects must be taken into
consideration has not been considered by us here. In 2D geometry, such an
analysis can be carried out using the methods of investigations of 2D
potential flows with a free boundary, which was proposed in \cite{28,29}
and is based on conformal mapping of the region occupied by the liquid to
a half-plane. In terms of the present work, such a transformation
corresponds to the use of the field potential $\varphi$ and its
harmonically conjugate function $v$ as independent variables. In the case
of the axial symmetry of the problem (such a geometry reflects the
experimentally observed phenomena \cite{12,24} most correctly), the
formation of singularities can be described by self-similar solutions of
the electrohydrodynamic equations, which are analogous to those considered
in the recent publication \cite{30} devoted to the formation of conic tips
on the surface of a liquid metal in an external electric field. In
accordance with the self-similar scenario of the development of
instability, conical dimples with an angle of $98.6^{\circ}$ appear on the
surface over a finite time. A detailed analysis of these processes calls
for further investigations. 

\smallskip

The author is grateful to V. E. Zakharov and E. A. Kuznetsov for their
interest in this research.


\begin{thebibliography}{}

\bibitem{1} M. W. Cole, M. H. Cohen, Phys. Rev. Lett. {\bf 23}, 1238
(1969).
  
\bibitem{2} V. B. Shikin, Zh. Eksp. Teor. Fiz., {\bf 58}, 1748 (1970)
[Sov.Phys. JETP {\bf 31}, 936 (1970)].

\bibitem{3} L.P. Gor'kov and D.M. Chernikova, Pis'ma Zh. Eksp. Teor. Fiz.,
{\bf 18}, 119 (1973) [JETP Lett. {\bf 18}, 68 (1973)].

\bibitem{4} D.M. Chernikova, Fiz. Nizk. Temp. {\bf 2}, 1374 (1976) [Sov.
J. Low Temp. Phys. {\bf 2}, 669 (1976)].

\bibitem{5} L. Tonks, Phys. Rev. {\bf 48}, 562 (1935). 

\bibitem{6} Ya.I. Frenkel', Zh. Eksp. Teor. Fiz., {\bf 6}, 347 (1936).

\bibitem{7} L.D. Landau and E.M. Lifshitz, {\sl Course of Theoretical
Physics, Vol.~8: Electrodynamics of Continuous Media}
(Nauka, Moscow, 1982; Pergamon, New York, 1984).

\bibitem{8} L.P. Gor'kov and D.M. Chernikova, Dokl. Akad. Nauk
SSSR, {\bf 228}, 829 (1976) [Sov. Phys. Dokl., {\bf 21}, 328
(1976)].

\bibitem{9} H. Ikezi, Phys. Rev. Lett. {\bf 42}, 1688 (1979).

\bibitem{10} D.M. Chernikova, Fiz. Nizk. Temp. {\bf 6}, 1513 (1980) [Sov.
J. Low Temp. Phys. {\bf 6}, 737 (1980)].

\bibitem{11} V. B. Shikin and Yu. P. Monarkha, {\sl Two-Dimensional Chared
Systems in Helium}, (Nauka, Moskow, 1989).

\bibitem{12} A. P. Volodin, M. S. Haikin, and V. S. Edel'man, Pis'ma Zh.
Eksp. Teor. Fiz. {\bf 26}, 707 (1977) [JETP Lett. {\bf 26}, 543 (1977)].

\bibitem{13} N. M. Zubarev, Pis'ma Zh. Eksp. Teor. Fiz. {\bf 71}, 534
(2000) [JETP Lett. {\bf 71}, 367 (2000)].

\bibitem{14} G. D. Crapper, J. Fluid Mech. {\bf 2}, 532 (1957).

\bibitem{15} V.E. Zakharov, Prikl. Mekh. Tekh. Fiz., No. 2, 86 (1968).

\bibitem{16} P. Ya. Polubarinova-Kochina, Dokl. Akad. Nauk SSSR, {\bf 47},
254 (1945).
 
\bibitem{17} D. Bensimon, L. P. Kadanoff, Sh. Liang, {\sl et al.,} Rev. Mod.
Phys. {\bf 58}, 977 (1986).

\bibitem{18} A. I. Dyachenko, V.E. Zakharov and E.A. Kuznetsov, Phys.
Plazmy {\bf 22}, 916 (1996).

\bibitem{19} V. E. Zakharov and A. I. Dyachenko, Physica D {\bf 98}, 652
(1996).

\bibitem{20} M. B. Mineev-Weinstein and S. P. Dawson, Phys. Rev. E {\bf
50}, R24 (1994).

\bibitem{21} N. M. Zubarev, Phys. Lett. A {\bf 243}, 128 (1998). 

\bibitem{22} N. M. Zubarev, Zh. Eksp. Teor. Fiz., {\bf 114}, 2043 (1998)
[Sov.Phys. JETP {\bf 87}, 1110 (1998)].

\bibitem{23} S. D. Howison, SIAM J. Appl. Math. {\bf 46}, 20 (1986). 

\bibitem{24} V. S. Edel'man, Usp. Fiz. Nauk, {\bf 130}, 675 (1980) 
[Sov. Phys. Usp., {\bf 23}, 227 (1980)].

\bibitem{25} G. B. Whitham, {\sl Linear and Nonlinear Waves}, (Wiley, New
York, 1974; Mir, Moskow, 1977).

\bibitem{26} N. M. Zubarev, Zh. Eksp. Teor. Fiz., {\bf 116}, 1990 (1999)
[JETP {\bf 89}, 1078 (1999)].

\bibitem{27} G. P. Ivantsov, Dokl. Akad. Nauk SSSR {\bf 58},
567 (1947).

\bibitem{28} A. I. Dyachenko, E. A. Kuznetsov, M. D. Spector, and V. E.
Zakharov, Phys. Lett. A {\bf 221}, 73 (1996).

\bibitem{29} A. I. Dyachenko, Dokl. Akad. Nauk {\bf 376}, 27 (2001).

\bibitem{30} N. M. Zubarev, Pis'ma Zh. Eksp. Teor. Fiz., {\bf 73}, 613
(2001) [JETP Lett. {\bf 73}, 544 (2001)].

\end{thebibliography}
\end{document}